\documentclass[aps,eqsecnum,nofootinbib,showpacs]{revtex4}
\usepackage[dvips]{graphicx}  
\usepackage{amsmath,amsfonts}
\usepackage{bm}
\usepackage{color}  
\begin{document}
\title{Weak gravitation from a small extra 2D sphere}
\author{Akira Kokado}
\email{kokado@kobe-kiu.ac.jp}
\affiliation{Department of Physical Therapy, Kobe International University, Kobe 658-0032, Japan}
\author{Takesi Saito}
\email{tsaito@k7.dion.ne.jp}
\affiliation{Department of Physics, Kwansei Gakuin University,
Sanda 669-1337, Japan}
\date{\today}
\begin{abstract}
 In order to explain weak gravitation in our 4-dimensional universe, a 6-dimensional model with a small extra 2D sphere is proposed. The traceless energy-momentum tensor is quite naturally appeared in our 6-dimensional model.
The warp factor is given by $\phi (\theta ) = \epsilon + \sin{\theta }$, where $\epsilon $ plays a role of killing the singular point $\phi (\theta )=0$, and is assumed $0 < \epsilon \ll 1$.  Any massive particle is rolling down into points   along this geodesic line. The light ray can be shown to stay in our 4-dimensional universe. This suggest us that our 4-dimensional world can be located at $\theta =0 $ and/or $\theta = \pi $, its background metric being  $\epsilon ^2 \eta _{\mu \nu }$. As a result, we have the 4-dimensional Newton constant, which is given by $G_N \simeq G_6 \epsilon ^{10}$ and the fifth force coefficients appeared here are  $\alpha _i\simeq \epsilon ^{2(i-4)}$, $i=1, 2, 3$. Here $G_6$ is the gravitational constant in 6-dimensional spacetime. If we take $\epsilon = 10^{-3.8}$ against $G_6\sim 1($GeV$)^{-2}$, we get $G_N\sim 10^{-38}($Gev$)^{-2}$, the present time gravitational constant.
\end{abstract}

\maketitle
\section{Introduction}\label{sec:intro}
%
%
\indent The gravitational force between two protons with mass $m$ is characterized by $G_N m^2 = m^2/M^2_P = 10^{-38}$, where $M_P$ is the Planck mass, while the electric force between them is characterized by  $e^2/\hbar c =1/137\simeq 10^{-2}$, $e$ being the electric charge of the proton. This extremely smallness of the gravitational force is widely known as one of the hierarchy problem. \\
\indent Why is the gravitational force so weak compared with the electromagnetic force?  This problem has been so far pursued by higher dimensional models \cite{ref:Randall,ref:Giddings, ref:Mannheim, ref:Garriga, ref:Tanaka, ref:Shiromizu, ref:Sasaki} rather than 4-dimensions, especially by the superstring theory \cite{ref:Green}, which requires 10-dimensional world volume. Comparing graviton with photon, graviton is made of closed string, while photon is made of open string, both ends of which are attached on the brane. So, closed string can freely move from D3-brane to extra dimensions, whereas open string cannot move to other dimensions, staying only on D3-brane. As a consequence the number of gravitons on the D3-brane is only decreasing, while the number of photons is not. This is the reason why the gravitational force becomes so weak compared with the electromagnetic force.  \\
\indent However, it may be so hard to solve exactly the hierarchy problem even by the superstring theory. A general overview is such that we have an equation 
for the 4-dimensional Newton constant $G_N$ 
\begin{align}
 & G_N = \frac{G_{4+d}}{R^{d}}~,
 \label{eq:Newton}
\end{align}
where $G_{4+d}$ is the gravitational constant in the (4+d)-dimensional spacetime, and $R$ is the size of compactified  dimensional extra-space. In this view the weak gravitation in our universe comes from so large $R$. \\ 
\indent Contrary to this, in the present paper we would like to propose a 6-dimensional model with the small extra- space with a background metric
\begin{align}
 & ds^2 = g_{IJ} dx^I dx^J
 \label{eq:definition_metric} \\
 & = \phi ^{2}(\theta ) \eta _{\mu \nu } dx^{\mu } dx^{\nu } + a^2 ( d\theta ^2 + \sin^2{\theta } d\varphi ^2 )~,
 \nonumber
\end{align}
where $x^{\mu }$ are 4-dimensional coordinates, while $ 0\leq  \theta \leq \pi$ and $0\leq \varphi \leq 2\pi $ are coordinates for extra two-dimensional spherical surface with a constant radius $a$, which is assumed to be very small. The warp factor $\phi (\theta )$ here is given by
\begin{align}
 & \phi (\theta ) = \epsilon + \sin{\theta }~, \quad (0< \epsilon \ll 1)
 \label{eq:warp_facter}
\end{align}
where the small parameter $\epsilon $ plays a role of that the warp factor does not vanish at $\theta =0, \pi $. We are amused to suppose that such a factor $\epsilon $ may come from something like quantum gravity effects to kill the singular point $\phi (\theta )=0$ \cite{ref:Bojowald}. \\
\indent We can easily see that the energy of any particle running along the geodesic line above is given by
\begin{align}
 & E = \Big( \frac{d\theta }{d\tau } \Big)^2  + \frac{c_6^{\ 2}}{\sin^2{\theta }} - \frac{A^2}{\phi ^2(\theta )}~,
 \label{eq:eq_of_motion}
\end{align}
where  $A^2=(c_0^{2}-\vec{c}^2)/a^2$, $dx^0/d\tau =c_0/\phi ^2$, $d\vec{x}/d\tau =\vec{c}/\phi ^2$, $d\varphi /d\tau =c_6/\sin^2{\theta }$  with $c_0$, $\vec{c}$, $c_6$ being  constants. When $c_6=0$,  the potential has a maximum value $-A^2$ at $\theta =\pi /2$ and the lowest value $-A^2/\epsilon ^2$  at $\theta =0, \pi $, so that any particle is rolling down into points $\theta =0, \pi $. As for massless photons we can see that they stay only in our 4-dimensional universe, but not traveling in extra-dimensions. This suggest us that our 4-dimensional world can be located at $\theta =0$ and/or $\theta =\pi $, and its background metric $g_{\mu \nu }$ becomes $\epsilon ^2 \eta _{\mu \nu }$.   \\
\indent For small fluctuations around this background we calculate the gravitational potential. In this formulation we would like to point out that the traceless energy-momentum tensor is quite naturally appeared in our 6-dimensional model. We owe this fact very much to put the traceless condition for weak gravitational field.. \\
\indent The gravitational potential related with test masses $M_0$ and $M_1$ in our 4-dimensional universe is calculated as
\begin{align}
 & V(r) = - G_N \frac{M_0 M_1}{r} \Big( 1 + \sum^{3}_{i=1} \alpha _i e^{-m_i r} \Big)~,
 \label{eq:potential}
\end{align}
where $G_N \simeq G_6 \epsilon ^{10}$ and $\alpha _i \simeq \epsilon ^{2(i-4)}$, $i=1, 2, 3$. Here $G_6$ is the gravitational constant in 6-dimensional spacetime with dimensions 2 in our model. The second exponential terms come from the KK modes which appear here only three as the fifth force, and rapidly drops off outside of the extra dimensions $a<r$. The hierarchy problem will be discussed in concluding remarks. \\
\indent In Sec.\ref{sec:2} the gravitational wave equation with the warp factor (\ref{eq:warp_facter}) is derived. In Sec.\ref{sec:3} we divide $\theta $ into three regions. We seek solutions of the wave equation in each region.  Connections of these solutions at boundaries are then considered. In Sec.\ref{sec:4} the gravitational potential in our 4-dimensional universe is calculated by means of the Green function. The final section is devoted to concluding remarks. The Appendix is prepared for calculations for the case of $\mu $ pure imaginary, where  $\mu $ is the suffix of the associated Legendre function. 
%
\section{Gravitational wave equation} \label{sec:2}
%
The Einstein equation with the 6-dimensional metric $g_{IJ}$ is given by 
\begin{align}
 &  R_{IJ} - \frac{1}{2}g_{IJ} R + \Lambda g_{IJ} = \kappa T_{IJ}~,
  \label{eq:Einstein_eq}
\end{align}
where $\kappa = 8\pi G_6$ and all others are in familiar notations. We rewrite this equation in the form
\begin{align}
 &  R_{IJ} - \frac{1}{2} g_{IJ} \Lambda = \kappa \Big( T_{IJ} - \frac{1}{4} g_{IJ} T \Big)~,
  \label{eq:Einstein_eq2} 
\end{align}
with $T = g^{IJ} T_{IJ}$.  The factor $1/4$ is characteristic in the 6-dimensional model. \\
\indent  We look for background solutions of Eq. (\ref{eq:Einstein_eq2}) for the warp factor $\phi (\theta )$ with the ansatz for stress-energy tensors of bulk matter fields \cite{ref:Gogberashvili, ref:Gogberashvili2, ref:Oda, ref:Kokado} 
\begin{align}
 & T_{\mu \nu } = g_{\mu \nu } f_1(\theta ) = \phi ^2(\theta ) \eta _{\mu \nu} f_1(\theta )~,
  \label{eq:Energy_momentum_tensor} \\
 & T_{55} = g_{55} f_2(\theta ) = a^2 f_2(\theta )~,
  \nonumber \\
 & T_{66} = g_{66} f_3(\theta ) = a^2 \sin^2{\theta } f_3(\theta )~.
  \nonumber
\end{align}
All other elements vanish. \\
\indent Relevant quantities are inserted into the ($\mu \nu $) component of Eq. (\ref{eq:Einstein_eq2}) to give

\begin{align}
 & 3 \frac{\phi'^2}{\phi ^2} + \frac{\phi''}{\phi } + \frac{\cos{\theta }}{\sin{\theta }}\frac{\phi'}{\phi }   + \frac{1}{2} \Lambda a^2 = \frac{1}{4}\kappa a^2 ( f_2 + f_3)~.
 \label{eq:equation_mu_nu_0}
\end{align}
In the same way, we get 
\begin{align}
 & 4\frac{\phi'' }{\phi } - 1 + \frac{1}{2} \Lambda a^2 = \frac{1}{4}\kappa a^2  (4f_1- 3f_2 + f_3)~,
 \label{eq:equation_55_0}
\end{align}
for the 55-component, and
\begin{align}
 & 4\frac{\cos{\theta }}{\sin{\theta }}\frac{\phi'}{\phi } - 1 + \frac{1}{2} \Lambda a^2 = \frac{1}{4}\kappa a^2  (4f_1 + f_2 - 3f_3)~,
 \label{eq:equation_66_0}
\end{align}
for the 66-component. From Eqs. (\ref{eq:equation_mu_nu_0})-(\ref{eq:equation_66_0}) we obtain
\begin{align}
 & f_1= \frac{1}{\kappa  a^2} \Big[ 3\frac{\phi ''}{\phi } + 3\frac{\phi '^2}{\phi ^2} + 3 \frac{\cos{\theta }}{\sin{\theta }}\frac{\phi '}{\phi } -1 + \Lambda a^2 \Big]~,
 \label{eq:value_f1} \\
 &  f_2 = \frac{1}{\kappa  a^2} \Big[ 6\frac{\phi '^2}{\phi ^2} + 4 \frac{\cos{\theta }}{\sin{\theta }}\frac{\phi '}{\phi } + \Lambda a^2 \Big] ~,
 \nonumber \\
  & f_3 = \frac{1}{\kappa  a^2} \Big[ 4\frac{\phi ''}{\phi } + 6\frac{\phi '^2}{\phi ^2}  + \Lambda a^2 \Big] ~.
  \nonumber
\end{align}
We assume that $f_1$, $f_2$ behave at most as $1/\sin{\theta }$ at $\theta \sim 0$, $\pi $. Under 
these boundary conditions we have a solution for $\phi (\theta )$ 
\begin{align}
 & \phi (\theta ) = \epsilon  + \sin{\theta }~, \quad (0< \epsilon \ll 1) 
 \label{eq:def_warp_facter}
\end{align}
The small parameter $\epsilon $ plays a role of killing singular point $\phi (\theta )=0$. We are amused to suppose that such a factor   may come from something like quantum gravity effects to kill the singular point at  $\phi (\theta )=0$ \cite{ref:Bojowald}.  The stress-energy tensor  $T_{IJ}$  is, therefore, given by substituting Eq. (\ref{eq:def_warp_facter}) into Eq. (\ref{eq:value_f1}). \\
\indent Let us now consider the most general perturbation $g_{IJ}^{(1)} = h_{IJ}$ around the background metric $g_{IJ}^{(0)}$, that is,
\begin{align}
 & ds^2 = g_{IJ} dx^I dx^J = (g_{IJ}^{(0)} + g_{IJ}^{(1)} )dx^{I} dx^{J} ~,
 \label{eq:definition_metric_a}
\end{align}
where
\begin{align}
 & g_{\mu \nu } = g^{(0)}_{\mu \nu }+g^{(1)}_{\mu \nu } = \phi ^2(\theta )\eta _{\mu \nu } + h_{\mu \nu } ~, \quad g^{(0)}_{\mu \nu } =  \phi ^2(\theta )\eta _{\mu \nu }~, \quad g^{(1)}_{\mu \nu }= h_{\mu \nu }~,
 \label{eq:metric_add_h} \\
 & g_{\mu 5} = g^{(1)}_{\mu 5} = h_{\mu  5}, \quad g_{\mu 6} = g^{(1)}_{\mu 6}= h_{\mu 6}~,
 \nonumber \\
 & g_{55} = g^{(0)}_{5 5}+g^{(1)}_{5 5 } = a^2 + h_{55}, \quad g_{66} = g^{(0)}_{6 6}+g^{(1)}_{6 6}=a^2 \sin^2{\theta } + h_{66}, \quad g_{56} = g_{56}^{(1)} = h_{56}~.
 \nonumber 
\end{align}
We put here the gauge conditions
\begin{align}
 & \partial _{I } h^{I}_{\ \mu } = 0~, \quad \partial _{I } h^{I}_{\ 5} = 0, \quad h_{5 6} =0~.
 \label{eq:gauge_condition}
\end{align}
\indent We now expand $R_{IJ}$, $T_{IJ}$ in order $h_{IJ}$ as
\begin{align}
 & R_{IJ} = R^{(0)}_{IJ} + R^{(1)}_{IJ}~, \quad T_{IJ} = T^{(0)}_{IJ} + T^{(1)}_{IJ}~.
 \label{eq:expand_h}
\end{align}
where
\begin{align}
 & T_{\mu \nu }^{(0)} = g_{\mu \nu }^{(0)} f_1 (\theta ) = \phi ^2 (\theta ) \eta _{\mu \nu } f_1(\theta )~,
 \label{eq:def_em_tensor_mn} \\
 & T_{55}^{(0)} =  g_{5 5}^{(0)} f_2 (\theta ) = a^2 f_2 (\theta )~,
 \nonumber \\
 & T_{66}^{(0)} =  g_{6 6}^{(0)} f_3 (\theta ) = a^2 \sin^2{\theta } f_3 (\theta )~.
 \nonumber
\end{align}
We shall restrict our attention to the most interesting case of a static particle with mass $M_0$ located at $\vec{x} = \theta = 0$. In this case the energy-momentum tensor is given by
\begin{align}
 & T_{\mu \nu }^{(1)} = \tau _{\mu \nu } (x)\delta (\theta) = M_0 \delta _{\mu } ^{0} \delta _{\nu } ^{0} \delta ^{(3)}(\vec{x}) \delta(\theta )~,
 \label{eq:em_tensor_static_mass} \\
 & T_{I J }^{(1)} = 0~, (I, J \neq 0)
 \nonumber
\end{align} 
$\tau _{\mu \nu}$ being in order $h_{\mu \nu }$. \\
\indent The first-order equation in Eq. (\ref{eq:Einstein_eq2} ) is given by
\begin{align} 
 & R_{IJ}^{\ (1)} - \frac{1}{2}g_{a b}^{\ (1)}\Lambda + \frac{1}{4}\kappa  g_{IJ}^{\ (1)}T^{(0)} + \frac{1}{4} \kappa g_{IJ}^{\ (0)} \tilde {T}^{(1)}= \kappa  \Big[T_{IJ}^{\ (1)}-\frac{1}{4} g_{IJ}^{\ (0)} T^{(1)}\Big] \equiv \kappa  \Sigma _{IJ}~.
 \label{eq:Einstein_Eq5_0}
\end{align}
where $T = g^{IJ} T_{IJ} = T^{(0)} + T^{(1)} + \tilde{T}^{(1)}$ with
\begin{align}
 & T^{(0)} = 4 f_1(\theta) + f_2(\theta) + f_3(\theta)~,
 \nonumber \\
 & T^{(1)} = g^{(0)\mu \nu } T_{\mu \nu}^{(1)} = \tau ^{\lambda }_{\lambda }(x) \delta (\theta )~,
 \nonumber \\
 & \tilde{T}^{(1)}= g^{(1)\mu \nu } T_{\mu \nu}^{(0)} = - h^{\lambda }_{\lambda } f_1(\theta) - h^{5}_{5} f_2(\theta ) - h^{6}_{6} f_3(\theta )~.
 \nonumber
\end{align}
The source terms $\Sigma _{IJ}$ are explicitly given by
\begin{align}
 & \Sigma _{\mu \nu }(x, \theta) = \Big[ \tau _{\mu \nu }(x) - \frac{1}{4} g_{\mu \nu }^{(0)}\tau (x) \Big]\delta (\theta) = \big( \delta _{\mu }^{\ 0}  \delta _{\nu }^{\ 0} + \frac{1}{4} \eta _{\mu \nu } \big) M_0 \delta^{(3)}(\vec{x}) \delta (\theta)~,
 \label{eq:source_munu} \\
 & \Sigma _{5 5}(x, \theta) = - a^2 \tau (x) \delta (\theta) = \frac{1}{4}\Big(\frac{a}{\epsilon }\Big)^2 M_0\delta^{(3)}(\vec{x}) \delta (\theta)~,
 \nonumber \\
 & \Sigma _{6 6}(x, \theta) = - \frac{1}{4} a^2 \sin^2{\theta} \tau (x) \delta (\theta) = 0~,
 \nonumber 
\end{align}  
where $\tau (x)\equiv \tau ^{\lambda }_{\ \lambda }(x)$. 
We solve Eq. (\ref{eq:Einstein_Eq5_0}) under such approximation that the source term $\Sigma _{5 5}$ is negligible compared with $\Sigma _{\mu \nu }$, that is, $|\Sigma _{5 5}/e_{a}^{2}| \ll |\Sigma _{\mu \nu }|$, where $a=a_0 e_a$, $e_a$ being the unit length. This will be realized by the assumption $(a_0/\epsilon ) \ll 1$. We can then put as $\Sigma _{5 5} \simeq 0$. \\
\indent The ($\mu \nu $) component equation of Eq. (\ref{eq:Einstein_Eq5_0}) is given by
\begin{align}
 & - \frac{1}{2}\Big[ \partial _I \partial^I h_{\mu \nu} + \partial _\mu \partial _\nu h^I_{\ I } \Big]
 \label{eq:munu_Einstein_eq3} \\
 & -\frac{1}{2 a^2}\Big[g_{\mu \nu }\frac{\phi '}{\phi }\partial _{\theta } h^{I }_{\ I } + \frac{\cos{\theta }}{\sin{\theta }}\partial _{\theta }h_{\mu \nu } - 8 \big( \frac{\phi ''}{\phi } +  \frac{{\phi '}^2}{\phi ^2} + \frac{\cos{\theta }}{\sin{\theta }}\frac{\phi '}{\phi }\big) h_{\mu \nu } + (2 - 2 \Lambda a^2 )h_{\mu \nu } \Big] 
 \nonumber \\
 & + \big(\frac{\phi '}{\phi } + \frac{1}{2}\frac{\cos{\theta }}{\sin{\theta }}\big) \big(\partial _\mu h^{5}_{\ \nu } + \partial _\nu h^{5}_{\ \mu } \big) 
 \nonumber  \\
 & + \big( \frac{3{\phi '}^2 + \phi  \phi ''}{\phi ^2} + \frac{\cos{\theta }}{\sin{\theta }}\frac{\phi '}{\phi }\big) h^{55} g_{\mu \nu }^{(0)} - \frac{\kappa }{4} g_{\mu \nu }^{(0)} \big( f_1 (\theta ) h^{\lambda }_{\ \lambda } + f_2(\theta ) h^{5}_{\ 5}  + f_3(\theta ) h^{6}_{\ 6} \big)
  \nonumber \\
 & = \kappa [ T^{(1)}_{\mu \nu } - \frac{1}{4} g_{\mu \nu }^{(0)} T^{(1)}] = \kappa \Sigma _{\mu \nu }(x, \theta)~.
 \nonumber
\end{align}
Taking the 4d-trace of this equation we have
\begin{align}
 & - \frac{1}{2}\Big[ \partial _I \partial^I h^{\lambda }_{\ \lambda } + \partial _\mu \partial^\mu h^I _{\ I} \Big]
 \label{eq:munu_Einstein_eq4} \\
 & - \frac{1}{a^2}\big(4\frac{\phi '}{\phi } + \frac{1}{2}\frac{\cos{\theta }}{\sin{\theta }}\big) \partial _{\theta }h^{\lambda }_{\ \lambda } - \frac{1}{a^2}\big(4\frac{\phi '}{\phi } + \frac{\cos{\theta }}{\sin{\theta }}\big) \partial _{\theta } h^{5}_{\ 5 } - \frac{2}{a^2}\frac{\phi '}{\phi } \partial _{\theta } h^{6}_{\ 6 }
 \nonumber  \\
 & + \frac{1}{a^2} \big( \frac{6{\phi '}^2 }{\phi ^2} + 4 \frac{\cos{\theta }}{\sin{\theta }}\frac{\phi '}{\phi } - \Lambda a^2 \big) h^{5}_{\ 5} - \frac{1}{a^2} \big( \frac{4\phi '' }{\phi } +  6\frac{{\phi '}^2 }{\phi ^2} + \Lambda a^2 \big) h^{6}_{\ 6} 
 = 0~.
 \nonumber
\end{align}
\indent The (55) component equation of Eq. (\ref{eq:Einstein_Eq5_0}) is given by
\begin{align}
 &  - \frac{1}{2} \big(  \partial _I  \partial ^I h_{5 5} + \partial _\theta^{\ 2} h^{I }_{\ I }\big)  - \frac{\phi '}{\phi } \partial _\theta h^{\lambda }_{\ \lambda } + \frac{1}{2} \frac{\cos{\theta }}{\sin{\theta }}\big( \partial _\theta h^{5}_{\ 5 } - 2 \partial _\theta h^{6}_{\ 6 } \big) 
  \label{eq:55_Einstein_eq4} \\ 
 & -\frac{1}{4}\big( 3\frac{\phi ''}{\phi } + 3\frac{(\phi ')^2}{\phi ^2} + 3 \frac{\cos{\theta }}{\sin{\theta }}\frac{\phi '}{\phi } -1 + \Lambda a^2 \big) h^{\lambda }_{\ \lambda } + \big( 4\frac{\phi ''}{\phi } +  \frac{9}{2}\frac{{\phi '}^2}{\phi ^2} + 3\frac{\cos{\theta }}{\sin{\theta }}\frac{\phi '}{\phi } +\frac{3}{4}\Lambda a^2 -1 \big)h^{5}_{\ 5}
 \nonumber \\
 & - \big( \frac{\phi ''}{\phi } + \frac{3}{2} \frac{(\phi ')^2}{\phi ^2}  + \frac{1}{4} \Lambda a^2 \Big) h^{6}_{\ 6}  = \kappa \Sigma _{5 5}(x, \theta) \simeq 0~.
 \nonumber
\end{align}
As special solutions of Eqs. (\ref{eq:munu_Einstein_eq4}) and (\ref{eq:55_Einstein_eq4}) we get approximately 
\begin{align}
 & h^{\lambda }_{\ \lambda } = h^{5}_{\ 5} = h^{6}_{\ 6} = 0~.
 \label{eq:special_solutions_munu}
\end{align}
\indent The (56) component of Eq.(\ref{eq:Einstein_Eq5_0}) is given by
\begin{align}
 &  \frac{1}{2}\big( \partial _\theta  \partial _I h^{I }_{\ 6} - \partial _\varphi  \partial _\theta  h^{I }_{\ I} \big)  + \frac{1}{2}\big(\frac{\phi '}{\phi } - \frac{\cos{\theta }}{\sin{\theta }} \big) \big( 2 \partial _\beta  h^{\beta }_{\ 6} - \partial _\varphi h^{\lambda }_{\lambda } \big)  + 2\frac{\phi '}{\phi } \partial _\varphi  h^{5}_{\ 5}= 0 ~. 
  \label{eq:def_Ricci_scalar_56_2}  
\end{align}
Hence, according to Eqs. (\ref{eq:gauge_condition}) and (\ref{eq:special_solutions_munu}) we get a special solution of this equation
\begin{align}
 & \partial _I h^{I}_{\ 6} = 0~.
 \label{eq:solution_56}
\end{align}
\indent As for the (66) component equations of Eq. (\ref{eq:Einstein_Eq5_0}), we get
\begin{align}
 &  \frac{1}{2} \big( 2\partial _\varphi  \partial _I h^{I}_{\ 6} - \partial _\varphi ^{\ 2} h^{I }_{\ I } - \partial _I \partial ^I h_{6 6}  \big) - \frac{1}{2} \sin{\theta }\cos{\theta }  \partial _{\theta } h^{I}_{I} - \big( 2\frac{\phi '}{\phi } - \frac{3}{2}\frac{\cos{\theta }}{\sin{\theta }}\big) \partial ^{\theta } h_{6 6}
 \label{eq:66_Einstein_eq3} \\
 & -\frac{1}{4} \sin^2{\theta} \big( 3\frac{\phi ''}{\phi } + 3\frac{(\phi ')^2}{\phi ^2} + 3 \frac{\cos{\theta }}{\sin{\theta }}\frac{\phi '}{\phi } -1 + \Lambda a^2 \big) h^{\lambda }_{\ \lambda } -\sin^2{\theta} \big( \frac{3}{2} \frac{\phi ^2}{\phi ^2} - 3 \sin{\theta }\cos{\theta }\frac{\phi '}{\phi } - \frac{1}{4} \Lambda a^2 -1 \big) h^{5}_{\ 5} 
 \nonumber \\
 &  + \sin^2{\theta}\big( 3\frac{\phi ''}{\phi } +  \frac{9}{2}\frac{{\phi '}^2}{\phi ^2} + 4\frac{\cos{\theta }}{\sin{\theta }}\frac{\phi '}{\phi } + \frac{3}{4}\Lambda a^2 -1-2 \frac{\cos^2{\theta}}{\sin^2{\theta}} \big)h^6_{\ 6}
 \nonumber \\
 & = \kappa \Sigma _{6 6}(x, \theta)= 0~,
 \nonumber
\end{align}
This equation is automatically satisfied by Eqs.(\ref{eq:special_solutions_munu}) and (\ref{eq:solution_56}). \\
\indent  For (5$\mu $) and (6$\mu $) components of Eq.(\ref{eq:Einstein_Eq5_0}) we get
\begin{align}
 & - \frac{1}{2}\big( \partial _I \partial ^I h_{\mu 5 } + \partial _\mu  \partial _\theta  h^{I }_{\ I }\big)
  \label{eq:def_Ricci_scalar_5mu2} \\
 & - \frac{1}{2} \big( \frac{\phi '}{\phi } - \frac{\cos{\theta }}{\sin{\theta }} \big) \big( 2 \partial _{\varphi } h^{6}_{\mu } - \partial _{\mu } h^{6}_{\ 6} \big) + \frac{1}{2}\big( 3\frac{\phi '}{\phi } + \frac{\cos{\theta }}{\sin{\theta }} \big) \partial _{\mu } h^{5}_{\ 5}
 \nonumber \\ 
 & -\big( 3\frac{{\phi '}^2}{\phi ^2} + \frac{\phi ''}{\phi } + \frac{\cos{\theta }}{\sin{\theta }}\frac{\phi '}{\phi } \big) h^{5}_{\ \mu } + \frac{1}{2a^2}\big( 8\frac{\phi ''}{\phi } +  12\frac{{\phi '}^2}{\phi ^2} + 8\frac{\cos{\theta }}{\sin{\theta }}\frac{\phi '}{\phi }\big)h_{5 \mu } + \frac{1}{a^2}\big( \Lambda a^2 -1) h_{5 \mu } = 0~.
  \nonumber 
\end{align}
and 
\begin{align}
 & \frac{1}{2}\big( \partial _\mu   \partial _I h^{I}_{\ 6 } - \partial _I \partial ^I h_{6 \mu } - \partial _\varphi  \partial _\mu  h^{I}_{\ I}\big)
  \label{eq:def_Ricci_scalar_6mu3} \\
 & + \big( 2 \frac{\phi '}{\phi } - \frac{1}{2} \frac{\cos{\theta }}{\sin{\theta }} \big)\partial _\varphi h^{5}_{\ \mu } -  \big( \frac{\phi '}{\phi } - \frac{1}{2} \frac{\cos{\theta }}{\sin{\theta }} \big)\partial ^{\theta }h_{6 \mu } - \big( \cos^2{\theta } + \sin{\theta }\cos{\theta } \frac{\phi '}{\phi } \big) h^{6}_{\ \mu }
  \nonumber \\
 & + \frac{1}{2a^2}\big( 8\frac{\phi ''}{\phi } +  12\frac{{\phi '}^2}{\phi ^2} + 8\frac{\cos{\theta }}{\sin{\theta }}\frac{\phi '}{\phi }\big)h_{6 \mu } + \frac{1}{a^2}\big( \Lambda a^2 -1) h_{6 \mu } = 0~.
  \nonumber
\end{align}
respectively. As special solutions of both equations we have
\begin{align}
 & h_{5 \mu } = h_{6 \mu} = 0~.
 \label{eq:solution_tmu_6mu}
\end{align}
according to results of Eqs. (\ref{eq:gauge_condition}), (\ref{eq:special_solutions_munu}) and (\ref{eq:solution_56}). \\
\indent Finally, thanks to results (\ref{eq:gauge_condition}), (\ref{eq:special_solutions_munu}), (\ref{eq:solution_56}) and (\ref{eq:solution_tmu_6mu}), Eq. (\ref{eq:munu_Einstein_eq3}) reduces to
\begin{align}
 & - \frac{1}{2\phi ^2} \Box h_{\mu \nu}  - \frac{1}{2 a^2 \sin^2{\theta }}\partial _\varphi^{\ 2} h_{\mu \nu} 
 \label{eq:munu_Einstein_eq_A} \\
 & -\frac{1}{2 a^2}\Big[\partial _\theta ^{\ 2} h_{\mu \nu } + \frac{\cos{\theta }}{\sin{\theta }}\partial _{\theta }h_{\mu \nu } - 8 \big( \frac{\phi ''}{\phi } +  \frac{{\phi '}^2}{\phi ^2} + \frac{\cos{\theta }}{\sin{\theta }}\frac{\phi '}{\phi }\big) h_{\mu \nu } + (2 - 2 \Lambda a^2 )h_{\mu \nu } \Big] 
 \nonumber \\
 & = \kappa [ \tau _{\mu \nu }(x) - \frac{1}{4} g_{\mu \nu }^{(0)} \tau (x)]\delta (\theta )~,
 \nonumber
\end{align}
where $\Box = \eta ^{\mu \nu }\partial _\mu \partial _\nu $. Note that we have imposed six gauge conditions (\ref{eq:gauge_condition}), but the other results (\ref{eq:special_solutions_munu}), (\ref{eq:solution_56}) and (\ref{eq:solution_tmu_6mu}) are not new gauge conditions, which are obtained as special solutions of Einstein equations in the gauges (\ref{eq:gauge_condition}).  The six gauge conditions are realized by six coordinate gauge functions $\varepsilon ^{I}(x)$. Namely, under an infinitesimal coordinate transformation  $\bar {x}^I = x^I + \varepsilon ^I(x)$, any general metric $g_{IJ}$ will transform in first order as $\bar{g}_{IJ} = g_{IJ} + \varepsilon _{I, J} + \varepsilon_{J, I}$. Since the number of gauge conditions are the same as that of $\varepsilon ^{I}$, we see that there occur no contradictions among six gauge conditions.
In the following we take a model of $\Lambda  a^2 = 3$. Substituting $\phi (\theta) = \epsilon  + \sin{\theta }$ into above, we get
\begin{align}
 & -\frac{1}{2\phi ^2} \Box h_{\mu \nu } -\frac{1}{2 a^2}\Big[ \partial _{\theta }^{\ 2}  + \frac{\cos{\theta }}{\sin{\theta }}\partial _{\theta }+20 - \frac{8+16\epsilon \sin{\theta }}{\phi \sin{\theta }}- \frac{8+16\epsilon \sin{\theta }+8\epsilon ^2}{\phi ^2} \Big]h_{\mu \nu }
\label{eq:f1} \\
 & - \frac{1}{2 a^2\sin^2{\theta }}\partial ^2_{\varphi } h_{\mu \nu } \equiv  L h_{\mu \nu } = \kappa  \Big[\tau _{\mu \nu }(x) -\frac{1}{4} \eta _{\mu \nu } \tau (x) \Big] \delta (\theta )~,
  \nonumber
\end{align}
\indent This equation can be solved by means of the Green function, which is given by eigenfunctions of the differential operator $L$ in the left-hand side, that is,
\begin{align}
 & L h_{\mu \nu } = 0~.
 \label{eq:def_L}
\end{align}
\indent Separating the four-dimensional mass term by
\begin{align}
 & h_{\mu \nu }(x, \theta , \varphi ) = h_{\mu \nu }(0) \exp{(i k_{\mu }x^{\mu })}  f(\theta ) k(\varphi )~, \quad k_{\mu } k^{\mu } = -m^2~,
 \label{eq:mass_condition}
\end{align}
with  $h^{\lambda }_{\ \lambda }(0)=0$, Eq. (\ref{eq:def_L}) becomes a simply separable equation. Hence we get $(\partial _{\varphi }^2 + n^2)k(\varphi )=0$, where $n$ should take integral values, because $k(\varphi )$ is a $2\pi $-periodic function. The equation for $f(\theta )$ is, therefore, given by 
\begin{align}
 & \Big[ \partial _{\theta }^{\ 2}  + \frac{\cos{\theta }}{\sin{\theta }}\partial _{\theta }+20 - \frac{n^2}{\sin^2{\theta }} - \frac{8+16\epsilon \sin{\theta }}{\sin{\theta }(\epsilon + \sin{\theta })}- \frac{8+16\epsilon \sin{\theta }+8\epsilon ^2-m^2a^2}{(\epsilon + \sin{\theta })^2} \Big]f(\theta ) = 0~.
 \label{eq:f_equation}
\end{align}
%
%
\section{Connection conditions} \label{sec:3}
%
%
\subsection{Three regions} \label{sec:3a}
%
Let us separate $\theta $ into three regions,
\begin{align}
 & \mbox{I.} \quad 0 \leq \theta \leq \alpha ~,
 \label{eq:regions} \\
 & \mbox{II.} \quad \alpha  \leq \theta \leq \pi - \alpha ~, 
 \nonumber \\
 & \mbox{III.} \quad \pi - \alpha   \leq \theta \leq \pi ~, 
 \nonumber
\end{align}
where $0<\epsilon \ll\alpha \ll 1$. We seek solutions of Eq. (\ref{eq:f_equation}) in each region. Then we consider connections of solutions at boundaries $\theta =\alpha $ and $\theta = \pi - \alpha $. \\
\indent In the region  II one can neglect $\epsilon $ in Eq. (\ref{eq:f_equation}), so that it reduces to

\begin{align}
 & \Big[ \partial _{\theta }^{\ 2}  + \frac{\cos{\theta }}{\sin{\theta }}\partial _{\theta }+\nu (\nu +1) - \frac{\mu ^2}{\sin^2{\theta }}  \Big]f(\theta ) = 0~,
 \label{eq:f_equation_1oeder}
\end{align}
where $\nu (\nu +1)=20$ and $\mu ^2=n^2 + 16 - m^2a^2$. This equation is nothing but the associated Legendre equation with $(\nu , \mu )=(4, \sqrt{n^2 +16-m^2a^2})$.  A general solution of Eq. (\ref{eq:f_equation_1oeder}) is, therefore, given by 
\begin{align}
 & f(\theta ) = A P_4^{\mu }(z) + B Q_4^{\mu }(z)~, \quad z = \cos{\theta }~.
 \label{eq:f_Legendre_fun} 
\end{align}
\indent In the region I we can use an approximation $\sin{\theta }\sim \theta $.  So, Eq. (\ref{eq:f_equation}) reduces to
\begin{align}
 & \Big[ \partial _{x}^{\ 2}  + \frac{1}{x}\partial _{x}+20 \epsilon ^2 - \frac{n^2}{x^2} + \frac{8}{x(1-x)}- \frac{8-m^2a^2}{(1-x)^2} \Big]f(x) = 0~,
 \label{eq:f_equation_x}
\end{align}
with $x=-\theta /\epsilon $. \\
\indent If we neglect $20\epsilon ^2$ and put $f(x)=(-x)^n (1-x)^{\delta }g(x)$, we get a solution for $g(x)$ in terms of Gaussfs hypergeometric function
\begin{align}
 & g(x) = F(\alpha , \beta , \gamma ; x)~,
 \label{eq:nu_vanish}
\end{align}
where 
\begin{align}
 & \alpha = n + \delta  + \mu ~, \quad \beta = n + \delta  - \mu ~, \quad \gamma = 2 n +1~,
  \quad \delta = \frac{1\pm\sqrt{33 - 4 m^2 a^2}}{2} ~.
  \label{eq:parameter1} 
\end{align}
Here it is enough to consider only the case of $\mu $ positive, because $F(\alpha , \beta , \gamma ; x)$  is symmetric under the exchange between $\alpha $ and $\beta $. \\
\indent Since $\gamma \in Z$, the general solution of $g(x)$ is given by \cite{ref:Abramowitz}
\begin{align}
 & g(x) = C_1 F(\alpha , \beta , \gamma ; x)
 \label{eq:solution_g} \\
 & + C_2 \Big[ F(\alpha , \beta , \gamma ; x) \ln{x} - \sum_{k=1}^{2n} \frac{(2n)_k (k-1)!}{(2n-k)!(1-\alpha )_k(1-\beta )_k}(-x)^{-k}
 \nonumber \\
 & + \sum_{k=0}^{\infty } \frac{(\alpha )_k (\beta )_k}{(2n+1)_k k!}x^k \big(\psi (\alpha +k) + \psi (\beta +k) - \psi (1+k) - \psi (2n+1+k)\big) \Big]~.
 \nonumber
\end{align}
In the following we do not interest the $n \neq 0$ solutions, because they do not contribute to the Green function $G_R(x_\mu , \theta =0;x_\mu ', \theta ' = 0)$. This can be seen from the equation $f(\theta ) = (\theta /\epsilon )^n (1+\theta /\epsilon )^\delta g(-\theta /\epsilon )$, which is zero at $\theta =0$ when $n \neq 0$, whereas not zero when $n=0$. So we put $n=0$ in Eq. (\ref{eq:solution_g}) to yield
\begin{align}
 & g(x) = C_1 F(\alpha , \beta , \gamma ; x)~.
 \label{eq:solution_g_n_0} \\
 & + C_2 \Big[ F(\alpha , \beta , \gamma ; x) \ln{x} + \sum_{k=0}^{\infty } \frac{(\alpha )_k (\beta )_k}{k!}x^k \big(\psi (\alpha +k) + \psi (\beta +k) - 2 \psi (1+k) \big) \Big]~.
 \nonumber
\end{align}
\indent Since $f(x)=(1-x)^\delta  g(x)$ should be finite at $x=0$, we must set always $C_2=0$. That is, the regular solution of Eq. (\ref{eq:solution_g_n_0}) with $n=0$ at $x=0$ is given by
\begin{align}
 & f(x) = C_1 (1-x)^{\delta } F(\alpha , \beta , \gamma ; x)~.
 \label{eq:solution_f_n_0_reguar}
\end{align}
\indent In the region III with notations, $\theta ' = \pi -\theta $, $x'= -\theta /\epsilon $, we have also the same regular solution at $x'=0$
\begin{align}
 & & f(x) = D_1 (1-x')^{\delta } F(\alpha , \beta , \gamma ; x')~.
 \label{eq:solution_f_n_0_reguar_pi}
\end{align}
%
\subsection{The case of $\mu $ integer} \label{sec:3b}
%
Around $\theta = \alpha $ one can use the formula in Ref.\cite{ref:Abramowitz2}.  For $\mu \neq0$ we have
\begin{align}
 & f(x) = C_1\Big[\frac{\Gamma (\delta +\mu )\sin{\pi (\delta -\mu )}}{\pi (2\mu )! \Gamma (\delta -\mu )}\ln{(-x)}(-x)^{-\mu } + \frac{\Gamma (2\mu )}{\Gamma (\delta +\mu )\Gamma (1-\delta +\mu )}(-x)^{\mu }
 \label{eq:case_inter_f_1order} \\
& + (-x)^{-\mu }\frac{\Gamma (\delta +\mu )\sin{\pi (\delta -\mu )}}{\pi (2\mu )! \Gamma (\delta -\mu )} \big(\psi (1) + \psi (2\mu +1) - \psi (1-\mu -\delta ) - \psi (\delta +\mu ) \big)\Big]~.
 \nonumber
\end{align}
Let us now consider connection conditions between several functions obtained above.
%
\subsubsection{$\mu =$1, 2, 3, 4} \label{sec:3b1}
%
This is the case of $m^2a^2=16 -\mu ^2$, $\delta =(1\pm\sqrt{4\mu ^2-31})/2$.  Around $\theta = \alpha $, Eq. (\ref{eq:case_inter_f_1order}) behaves as

\begin{align}
 & f(x) \simeq  C_1 \big(\frac{\theta }{\epsilon }\big)^{\mu } d_{\mu }~,
  \label{eq:f_mu=4}
\end{align}
with 
\begin{align}
 & d_{\mu } = \frac{\Gamma (2\mu )}{\Gamma (\delta +\mu )\Gamma (\mu + 1 -\delta )}~.
 \label{eq:def_d_mu}
\end{align}
The equation (\ref{eq:f_mu=4}) is compared with the solution (\ref{eq:f_Legendre_fun}) in the same region, i.e.,

\begin{align}
  & f(z) = A_{\mu } P_4^{\mu }(\cos{\theta }) + B_{\mu } Q_4^{\mu }(\cos{\theta }) \sim A_{\mu } \beta _{\mu } \theta ^\mu  + B\frac{{\beta '}_{\mu }}{\theta ^\mu }~,
 \label{eq:solution_Legendre_4} 
\end{align}
where
\begin{align}
 & \beta _1 = 10~, \quad \beta _2 = 45~, \quad \beta _3 = 105~, \quad \beta _4 = 10~,
 \label{eq:value_beta} \\
 & {\beta '}_1 = 1~, \quad {\beta '}_2 = 2~, \quad {\beta '}_3 = 8~, \quad {\beta '}_4 = 48~,
 \nonumber
\end{align}
and  
\begin{align}
 & A_{\mu }=C_1 \frac{d_{\mu }}{{\beta '}_{\mu }\epsilon ^{\mu }}~, \quad B_{\mu }=0~.
 \label{eq:relation_A_B_C1}
\end{align}
\indent The same connection should be taken at $\theta = \pi - \alpha $.  Around $\theta ' = \alpha $,  the regular solution with notations, $\theta '=\pi -\theta $,  $x'=-\theta '/\epsilon $, is given by
\begin{align}
 & f(x') \simeq  D_1 \big(\frac{\theta '}{\epsilon }\big)^\mu d_{\mu }~.
  \label{eq:f_mu=4_pi}
\end{align}
This is compared with the solution (\ref{eq:f_Legendre_fun}) in the same region, i.e.,
\begin{align}
 & f(\cos{\theta }) = f(\cos{(\pi -\theta ')}) \sim (-1)^{\mu } A_{\mu }\beta _{\mu } {\theta '}^\mu  + (-1)^{\mu +1} B_{\mu }\frac{{\beta '}_{\mu }}{{\theta '}^{\mu }}~,
 \label{eq:solution_Legendre_4_pi}
\end{align} 
to get  $A_{\mu }=(-1)^{\mu } D_1 \frac{d_{\mu }}{\beta _{\mu }\epsilon ^{\mu }}$, $B_{\mu }=0$. Since $f(\theta )=A_{\mu } P_4^{\mu }(\cos{\theta })$ is (anti-)symmetric under the exchange between $\theta $ and $\theta '$, we have $D_1=(-1)^{\mu }C_1$. \\
\indent To sum up, we have solutions with $\mu =$1, 2, 3 and 4, i.e., 
\begin{align}
 f(\theta ) = \begin{cases}
				A_{\mu } \frac{\beta _{\mu }}{d_{\mu }}\epsilon ^{\mu } \big(1+ \frac{\theta }{\epsilon }\big)^{\delta }F\big(\delta +\mu , \delta -\mu , 1 ; -\frac{\theta }{\epsilon }\big) ~, &  (0\leq \theta \leq  \alpha ) \\
				A_{\mu } P_4^{\mu }(\cos{\theta })~, &  (\alpha \leq \theta \leq \pi - \alpha ) \\
				(-1)^{\mu }A_{\mu } \frac{\beta _{\mu }}{d_{\mu }}\epsilon ^{\mu } \big(1+ \frac{\pi -\theta }{\epsilon }\big)^{\delta }F\big(\delta +\mu , \delta -\mu , 1 ; -\frac{\pi -\theta }{\epsilon }\big)~. &  (\pi -\alpha \leq \theta \leq  \pi ) 
			\end{cases}
 \label{eq:solution_1234}
\end{align}
\indent Normalizations for solutions with $\mu =$1, 2, 3, 4 are given by
\begin{align}
 & 1 = \int_{-1}^{1} dz~ |f(z)|^2 \simeq \int_{-1}^{1} dz~  |A_{\mu } P_{4}^{\mu }(z)|^2 = |A_{\mu }|^2 \frac{2}{9}\frac{(4+\mu )!}{(4-\mu )!}~,
 \label{eq:normalzations_f} \\
 &  |A_{\mu }|^2 = \frac{9}{2}\frac{(4-\mu )!}{(4+\mu )!}~.
 \nonumber
\end{align}
Here we have neglected small contributions which come from small regions I and III.
%
%
%
\subsubsection{$\mu =0$} \label{sec:3b2}
%
%
This is the case of $m^2a^2=16$, $\delta = \frac{1\pm i\sqrt{31}}{2}$.  Around $\theta = \alpha $, the 
formula with $\mu=0$ in Ref.\cite{ref:Abramowitz2} behaves like
\begin{align}
 & f(x) = C_1 \Big[\frac{\sin{\pi \delta }}{\pi }\ln{\frac{\theta }{\epsilon }} + \frac{\sin{\pi \delta }}{\pi }\big( 2\psi (1) - \psi (1-\delta ) - \psi (\delta )\big) \Big]~.
 \label{eq:f_mu_0}
\end{align}
Note that the second term in the formula in Ref.\cite{ref:Abramowitz2} does not appear for the $\mu =0$ case. The equation (\ref{eq:f_mu_0}) is compared with the solution (\ref{eq:f_Legendre_fun}.2) in the same region, i.e.,
\begin{align}
  & f(z) = A P_4(\cos{\theta }) + B Q_4(\cos{\theta }) \sim A   + B \ln{\frac{2}{\theta }}= B \ln{\frac{\epsilon }{\theta }} + A + B \ln{\frac{2}{\epsilon }}~,
 \label{eq:solution_Legendre_0} 
\end{align}
to give
\begin{align}
 & B= - C_1 \frac{\sin{\pi \delta }}{\pi }~,
  \label{eq:relation_B_C_1} \\
 & A= -B \Big[ \ln{\frac{2}{\epsilon }} + 2\psi (1) - \psi (1-\delta ) - \psi (\delta ) \Big]~.
  \label{eq:relation_A_B}
\end{align}
\indent The same connection should be taken at $\theta = \pi -\alpha $.  Around $\theta = \alpha $, Eq. (\ref{eq:case_inter_f_1order}) is given by
\begin{align}
 & f(x') = D_1  \frac{\sin{\pi \mu }}{\pi }\Big[ \ln{\frac{\theta '}{\epsilon }} + 2\psi (1) - \psi (1-\delta ) - \psi (\delta )\Big]~.
 \label{eq:f_pi}
\end{align}
This is compared with the solution (\ref{eq:f_Legendre_fun}) in the same region, i.e.,
\begin{align}
 & f(\cos{\theta }) = f(\cos{(\pi -\theta ')}) \sim A  - B\ln{\frac{2}{\theta '}}~,
 \label{eq:solution_Legendre_0_pi}
\end{align} 
to give
\begin{align}
 & A= B \Big[ \ln{\frac{2}{\epsilon }} + 2\psi (1) - \psi (1-\delta ) - \psi (\delta ) \Big]~.
 \label{eq:relation_A_B_2}
\end{align}
The plus sign of $A$ contradicts with the minus sign of Eq. (\ref{eq:relation_A_B}). So we conclude that there is no solution with $\mu =0$.
%
%
\subsection{The case of $\mu \notin Z$} \label{sec:3c}
%
In this section we consider the case of $\mu \notin Z$. We would like to show that there are no solutions with such $\mu $ real. \\
\indent It is enough to consider only a case of $\mu $ real positive, as already noted before. In this case the equation
\begin{align}
 & f(z) = A_{\mu } P_4^{\mu }(z) + B_{\mu } Q_4^{\mu }(z)~, \quad z=\cos{\theta }~, \quad \alpha \leq \theta \leq \pi -\alpha ~,
 \label{eq:Legendre_1oder}
\end{align}
can be written in the following form:
\begin{align}
 & f(z) = C_{\mu } \Big[\frac{1+z}{1-z}\Big]^{\mu /2} f_\mu (z) + D_{\mu } \Big[\frac{1+z}{1-z}\Big]^{-\mu /2} f_{-\mu }(z) ~,
 \label{eq:def_f_mu}
\end{align}
where
\begin{align}
 & f_{\mu}(z) = F(5, -4, 1-\mu , \frac{1-z}{2})
 \label{eq:def_f_mu2} \\
 & = 1 - \frac{10(1-z)}{1-\mu} + \frac{45(1-z)^2}{(1-\mu)(2-\mu)} -\frac{105(1-z)^3}{(1-\mu)(2-\mu)(3-\mu)} + \frac{105(1-z)^4}{(1-\mu)(2-\mu)(3-\mu)(4-\mu)}~.
 \nonumber
\end{align}
Around $\theta = \alpha $ we still take Eq. (\ref{eq:solution_f_n_0_reguar}), i.e.
\begin{align}
 & f(x) = C_1 (1-x)^{\delta } F(\alpha , \beta , \gamma ; x)~.
 \label{eq:f_mu_1order}
\end{align}
In this region, one can use the formula\cite{ref:Abramowitz3} to yield
\begin{align}
 & f(\theta ) \simeq C_1 \Big[ \frac{\Gamma (2\mu)}{\Gamma(1-\delta +\mu )\Gamma(\delta +\mu)}\Big(\frac{\theta}{\epsilon }\Big)^{\mu} + \frac{\Gamma (-2\mu)}{\Gamma(1-\delta - \mu )\Gamma(\delta - \mu)}\Big(\frac{\theta}{\epsilon }\Big)^{-\mu} \Big]~.
 \label{eq:F_1order}
\end{align}
This is compared with Eq. (\ref{eq:def_f_mu}) around $\theta = \alpha$,  i.e.,
\begin{align}
 & f(x) = C_{\mu } \Big(\frac{2}{\theta }\Big)^\mu  + D_{\mu } \Big(\frac{2}{\theta }\Big)^{-\mu}~,
 \label{eq:Legendre_mu_1order}
\end{align}
to give
\begin{align}
 & C_{\mu }= C_1 \frac{\Gamma (-2\mu )}{\Gamma (1-\delta -\mu )\Gamma (\delta -\mu )}\Big( \frac{\epsilon }{2}\Big)^{\mu } \simeq 0 ~,
 \label{eq:relation_C_C1} \\
 & D_{\mu }= C_1 \frac{\Gamma (2\mu )}{\Gamma (1-\delta +\mu )\Gamma (\delta +\mu )}\Big( \frac{2}{\epsilon }\Big)^{\mu } ~.
 \label{eq:relation_D_C1}
\end{align}
These equations tell us $C_{\mu }\simeq 0$ around $\theta = \alpha $, when $\mu $ is real positive. \\
\indent Next we consider the connection at $\theta = \pi - \alpha $.   Around $\theta '=\alpha $,  the regular solution with notations, $\theta ' = \pi - \theta $, $x'=-\theta '/\epsilon $, is given by
\begin{align}
 & f(x') = D_1 (1-x')^{\delta } F(\alpha , \beta , \gamma ; x')
 \label{eq:f_mu_1order_pi} \\
 & \simeq  D_1 \Big[ \frac{\Gamma (2\mu)}{\Gamma(1-\delta +\mu )\Gamma(\delta + \mu)}\Big(\frac{\theta'}{\epsilon }\Big)^{\mu} + \frac{\Gamma (-2\mu)}{\Gamma(1-\delta - \mu )\Gamma(\delta -\mu)}\Big(\frac{\theta'}{\epsilon }\Big)^{-\mu}\Big]~.
 \nonumber
\end{align}
On the other hand, Eq. (\ref{eq:def_f_mu}) behaves as, in the same region,
\begin{align}
 & f(\cos{\theta }) =  f(\cos{(\pi -\theta ')}) \simeq  C_{\mu } \Big(\frac{\theta '}{2}\Big)^{\mu } f_\mu (-1) + D_{\mu } \Big(\frac{\theta '}{2}\Big)^{-\mu } f_{-\mu} (-1) ~.
 \label{eq:f_mu_pi}
\end{align}
Comparing Eq.(\ref{eq:f_mu_pi}) with Eq.(\ref{eq:f_mu_1order_pi}) we have
\begin{align}
 & C_{\mu } = D_1 \frac{\Gamma (2\mu)}{\Gamma(1-\delta +\mu )\Gamma(\delta +\mu)f_{\mu}(-1)}\Big(\frac{2}{\epsilon }\Big)^{\mu} ~,
 \label{eq:relation_C_D1} \\
& D_{\mu } = D_1 \frac{\Gamma (-2\mu)}{\Gamma(1-\delta - \mu )\Gamma(\delta -\mu)f_{-\mu}(-1)}\Big(\frac{\epsilon }{2}\Big)^{\mu} \simeq 0~.
 \label{eq:relation_D_D1}
\end{align}
Both results $C_{\mu }=D_{\mu } \simeq 0$ in Eqs. (\ref{eq:relation_C_C1}) and (\ref{eq:relation_D_D1}) mean that there are no solutions with $\mu $ real positive other than  1, 2, 3, 4.
%
%
\subsection{The case of $\mu$ pure imaginary} \label{sec:3d}
%
%
We set $\mu =i \mu '$ with $\mu '$ real. Note that the case of $\mu '=0$ is ruled out by the definition of Eq. (\ref{eq:def_f_mu}). From Eqs. (\ref{eq:relation_C_C1}) and (\ref{eq:relation_D_C1}) with $\mu = i \mu '$ it follows that
\begin{align}
 & \frac{C}{D} = \frac{\Gamma (-2i\mu ') \Gamma (1-\delta +i\mu' )\Gamma (\delta +i\mu ')}{\Gamma (2i\mu ') \Gamma (1-\delta -i\mu ')\Gamma (\delta -i\mu ')}\Big(\frac{\epsilon }{2}\Big)^{2i\mu'}~,
 \label{eq:relation_C_D_imaginary}
\end{align}
while from Eqs. (\ref{eq:relation_C_D1}) and (\ref{eq:relation_D_D1})
\begin{align}
 & \frac{C}{D} = \frac{\Gamma (2i\mu ') \Gamma (1-\delta -i\mu ')\Gamma (\delta -i\mu ')}{\Gamma (-2i\mu ') \Gamma (1-\delta +i\mu ')\Gamma (\delta +i\mu ')}\frac{f_{-i\mu '}(-1)}{f_{i\mu '}(-1)}\Big(\frac{\epsilon }{2}\Big)^{2i\mu'}~.
 \label{eq:relation_C_D_imaginary2}
\end{align}
Both equations (\ref{eq:relation_C_D_imaginary}) and (\ref{eq:relation_C_D_imaginary2}) yield
\begin{align}
 & \Big(\frac{\epsilon }{2}\Big)^{4i\mu '} = \Big[\frac{\Gamma (2i\mu ') \Gamma (1-\delta -i\mu ')\Gamma (\delta -i\mu ')(1-i\mu ')(2-i\mu ')(3-i\mu ')(4-i\mu ')}{\Gamma (-2i\mu ') \Gamma (1-\delta +i\mu ')\Gamma (\delta +i\mu ')(1+i\mu ')(2+i\mu ')(3+i\mu ')(4+i\mu ')}\Big]^2~.
 \label{eq:condition_mu}
\end{align}
In the Appendix , however, it is shown that there are no solutions for $\mu '$ in this equation
%
\section{Green function} \label{sec:4} 
%
The Einstein equation for $h_{\mu \nu }$ is given by Eq. (\ref{eq:f1}) in 6-dimensions. However, we have set $n=0$ for eigenfunctions $k(\varphi )$ in Eq.(\ref{eq:mass_condition}), so that the  $\varphi $-dependence completely disappears from the theory. So, let us write Eq.(\ref{eq:f1}) in the 5-dimensional form:
\begin{align}
 & L h_{\mu \nu } = \kappa \Sigma_{\mu \nu } (x, \theta )~,
 \label{eq:f1_4}
\end{align}
where
\begin{align}
 & L = -\frac{1}{2\phi ^2(\theta )}\partial _{\lambda } \partial^{\lambda } - \frac{1}{2a^2}\big[ \partial_{\theta }^2 + \frac{\cos{\theta }}{\sin{\theta }}\partial _{\theta } + 20 - \frac{8}{\phi ^2(\theta )} - \frac{8}{\phi (\theta ) \sin{\theta }}\big]~,
 \label{eq:Laplacian_4}
\end{align}
\begin{align}
 & \Sigma _{\mu \nu }(x, \theta ) = \big( \tau _{\mu \nu }(x) - \frac{1}{4} \eta _{\mu \nu }\tau (x)\big) \delta (\theta )~.
 \label{eq:Laplacian_4a}
\end{align}
Here $\Sigma_{\mu \nu } (x, \theta )$  is the traceless energy-momentum tensor, consistent with the traceless condition $h^{\mu }_{\mu }=0$. We shall restrict our attention to the most interesting case of a static particle with mass $M_0$ located at $\vec{x}=\theta =0$. In this case the energy-momentum tensor is given by $\tau _{00}(x) = M_0 \delta (\vec{x})$ and others = 0. Then we have $\Sigma _{00}(X) = (3/4)M_0 \delta (\vec{x}) \delta (\theta )$.  \\ 
\indent The solution $h_{\mu \nu }$ can be derived by means of the Green function as
\begin{align}
 & h_{\mu \nu }(X) = \int d^5 X'~G_R(X, X') \kappa \Sigma_{\mu \nu } (X'), \quad X=(x, \theta )~,
 \label{eq:h_Green_fun}
\end{align}
The Green function is defined by
\begin{align}
 & L G_R(X, X') = \delta ^5 (X - X')~,
 \label{eq:def_Green_fun}
\end{align}
where
\begin{align}
 & G_R(X, X') = L^{-1} \sum_m h_m(X) h_m^{\dagger}(X')
 \label{eq:Cal_Green1} \\
 & = (-2) \sum_m \int \frac{d^4 p}{(2\pi )^4} \frac{e^{i p(x-x')}}{\frac{1}{\phi ^2(\theta )}(-p^2-m^2)}h_m(\theta ) h_m^{\dagger}(\theta ')~.
 \nonumber
\end{align}
Here $h_m(\theta )$  are mass eigenfunctions of the differential operator $L$. \\
\indent Following Tanaka et al. [4] we define the stationary Green function by 
\begin{align}
 & G_R(\vec{x}, \theta ;\vec{x'}, \theta ') = \int _{-\infty }^{\infty } dt'  G_R(X, X')
 \label{eq:Cal_Green2} \\
 & = 2 \sum_m \int \frac{d^3 p}{(2\pi )^3} \frac{e^{i \vec{p}(\vec{x}-\vec{x}')}}{\frac{1}{\phi ^2(\theta )}(\vec{p}^2+m^2)}h_m(\theta ) h_m^{\dagger}(\theta ')
 \nonumber \\
 & = 2 \phi ^2(\theta ) \frac{1}{4\pi r}\Big[ h_0(\theta ) h_0^{\dagger}(\theta ) + \sum_{m \neq 0} e^{-mr}h_m(\theta ) h_m^{\dagger}(\theta ') \Big]~,
 \nonumber
\end{align}
where $r = |\vec{x} - \vec{x'}|$.  The gravitational potential between masses $M_0$ and $M_1$   separated by $r$ is given, by putting $\theta =\theta '= \vec{x'} =0$, as follows:
\begin{align}
 & V(r) = - \frac{1}{2} M_{1} h_{00}(r) = -\frac{3}{8} M_0 M_1 G_R(\vec{x}, 0; \vec{0}, 0)
 \label{eq:Cal_Green_fun3} \\
 & = - 6\pi G_{6} M_0 M_1 \phi ^2(0) \frac{1}{4\pi r} \Big[ h_0(0) h_0^{\dagger}(0) + \sum_{m \neq 0} e^{-m r} h_m(0) h_m^{\dagger}(0) \Big]~,
 \nonumber 
\end{align}
where mass eigenstates are given by Eqs. (\ref{eq:solution_1234}), 
%
%
hence 
\begin{align}
 & h_m(0) = A_{\mu } \frac{\beta _{\mu }}{d_{\mu }}\epsilon ^{\mu } \equiv c_{\mu } \epsilon ^{\mu } \quad \mu = \sqrt{16-m^2 a^2}
 \label{eq:mass_eignstates}
\end{align}
with
\begin{align}
 m = \begin{cases}
				0 ~, &  (\mu =4) \\
				\frac{\sqrt{7}}{a}~, &  (\mu =3) \\
				\frac{\sqrt{12}}{a}~, &  (\mu =2) \\
				\frac{\sqrt{15}}{a}~. &  (\mu =1) \\
			\end{cases}
 \label{eq:valu_mass}
\end{align}
\indent Finally we have
\begin{align}
 & V(r) = - G_N \frac{M_0 M_1}{r} \Big( 1 + \sum_{i=1}^{3} \alpha _{i} e^{-m r_i} \Big)~,
 \label{eq:gravitational potential}
\end{align}
with
\begin{align}
 & G_N = 6 c_4^2 \epsilon ^{10} G_6~,
  \label{eq:defin_GN} \\
 & \alpha _{i} = \Big(\frac{c_i}{c_4}\Big)^2 \epsilon ^{2(i-4)}~, \quad (i=1, 2, 3)
 \nonumber
\end{align}
\indent The gravitational potential for another point at $\theta =\pi $ is also calculated in the same way as above. However, we have the same result as above, because mass eigenfunctions at $\theta =\pi $ are given by Eqs. (\ref{eq:solution_1234}), and their absolute values are irrelevant to the sign factor $(-1)^{\mu }$. \\
\indent Note that the 6-dimensional gravitational constant $G_6$ has the same dimensions 2 as that of $G_N$ in our model, because $\theta , \varphi $ have no dimensions.
%
%
\section{Concluding remarks} 
%
\indent We have proposed the 6-dimensional model with the metric (\ref{eq:definition_metric}), in order to explain weak gravitation from small extra dimensions $a/e_a \ll 1$, $e_a$ is the unit length.  The traceless energy-momentum tensor is quite naturally obtained in our 6-dimensional model. We owe this fact very much to put the traceless gauge condition for weak gravitational field. \\
\indent The warp factor is given by $\phi (\theta )=\epsilon +\sin{\theta }$, where $\epsilon $ plays a role of killing the singular point $\phi (\theta )=0$, and is assumed $\epsilon \ll 1$. As noted in the introduction, we are enough to calculate gravitational potentials for sources at $\theta =0$ and/or at $\theta =\pi $, though both potentials happen to be completely the same. \\
\indent We have the gravitational potential for test masses $M_0$ and $M_1$ in our 4-dimensional universe, which is given by Eq. (\ref{eq:gravitational potential}). Here important results are $G_N\simeq G_6\epsilon ^{10}$ and $\alpha _{i}\simeq \epsilon ^{2(i-4)}$, $i=1, 2, 3$. The second exponential terms come from the KK modes which appear here only three as the fifth forces, and rapidly drops off outside of the extra dimensions $a<r$. If we take $\epsilon =10^{-3.8}$ against $G_6\sim 1$(GeV$)^{-2}$, we get $G_N\sim 10^{-38}$(GeV$)^{-2}$, the present time gravitational constant. \\
\indent Inside of the extra dimensions $a>r$, the second exponential terms dominate. It is larger than 1 by at most $\alpha _i \sim 10^{23}$, so that the effective Newton constant becomes  $G_N\times 10^{23}\simeq 10^{-15}$(GeV$)^{-2}$. This value is still small but extremely larger than $G_{N}=10^{-38}($GeV$)$ by $10^{23}$. If we succeed experimentally to check such large values of the fifth force coefficients $\alpha _i$, we then conclude that there exists certainly the extra dimensional world. \\
\indent The reason why the gravitational force is so weak compared with the electromagnetic force is that the gravitational wave is going out of our 4-dimensional universe and running into the small extra 2D sphere as the KK modes, in which their effects are given by large enough value of $\alpha_i$. On the other hand massless photons stay in our 4D universe, keeping the same electromagnetic strength.\\
\indent We think that the so-called hierarchy problem appears here in something like a weaker form. \\
\section*{Acknowledgments}

We would like to express our sincere thanks to T. Okamura for many helpful discussions.
\appendix
\section{The case of $\mu $  pure imaginary}\label{sec:appendA}
%
\indent We consider the equation (3.43), i.e.,
\begin{align}
 & \Big(\frac{\epsilon }{2}\Big)^{4i\mu '} = \Big[\frac{\Gamma (2i\mu ') \Gamma (1-\delta -i\mu' )\Gamma (\delta -i\mu' )(1-i\mu ')(2-i\mu ')(3-i\mu ')(4-i\mu ')}{\Gamma (-2i\mu ') \Gamma (1-\delta +i\mu ')\Gamma (\delta +i\mu ')(1+i\mu ')(2+i\mu ')(3+i\mu ')(4+i\mu ')}\Big]^2~,
 \label{eq:condition_mu2}
\end{align}
In the following it is shown that this equation has no solutions for $\mu '$. The case of $\mu '=0$ is ruled out by the definition. \\
\indent We use $\delta =1/2 + i \sigma $, $\mu ' = m^2 a^2 - 16 >0$, $\sigma ^2 = m^2 a^2 -33/4>0$. Hence (\ref{eq:condition_mu2}) can be written as
\begin{align}
 & \Big(\frac{\epsilon }{2}\Big)^{4i\mu '} = \Big[\frac{\Gamma (2i\mu ') \Gamma (\frac{1}{2}-i(\mu '+ \sigma ))\Gamma (\frac{1}{2}- i(\mu ' - \sigma ))(1-i\mu ')(2-i\mu ')(3-i\mu ')(4-i\mu ')}{\Gamma (-2i\mu ') \Gamma (\frac{1}{2}+i(\mu ' + \sigma ))\Gamma (\frac{1}{2} +i(\mu '- \sigma ))(1+i\mu ')(2+i\mu ')(3+i\mu ')(4+i\mu ')}\Big]^2~,
 \label{eq:condition_mu2a}
\end{align}
 According to formulaes
\begin{align}
 & \Gamma (z) = \frac{1}{z} \prod_{m=1}^{\infty } \frac{\big(1+\frac{1}{m}\big)^z}{1+\frac{z}{m}}~,
 \label{eq:formura_Gamma} \\
 & \ln{\frac{1 + i x}{1 - i x}} = 2 i \tan^{-1}{x}~,
 \label{eq:formura_ln}
\end{align}
Eq. (\ref{eq:condition_mu2a}) reduces to
\begin{align}
 & \mu ' \ln{\big(\frac{\epsilon }{2}\big)} = \tan^{-1}{2(\mu ' + \sigma )} +  \tan^{-1}{2(\mu ' - \sigma )}
  \label{eq:cal_log_0} \\ 
 & +  \tan^{-1}{(-\mu ')} +  \tan^{-1}{\frac{(-\mu ')}{2}} +  \tan^{-1}{\frac{(-\mu ')}{3}} +  \tan^{-1}{\frac{(-\mu ')}{4}}
 \nonumber \\
 & + \sum_{m=1}^{\infty }\big( \tan^{-1}{\frac{(-2\mu ')}{m}} + \tan^{-1}{\frac{\mu ' + \sigma }{m  + \frac{1}{2}}} + \tan^{-1}{\frac{\mu ' - \sigma }{m  + \frac{1}{2}}}\big) ~.
 \nonumber
\end{align}
Thanks to the formula
\begin{align}
 & \tan{\big(\alpha +\beta +\gamma \big)} = \frac{ \tan{\alpha } + \tan{\beta } + \tan{\gamma } -  \tan{\alpha } \tan{\beta } \tan{\gamma }}{ 1- \tan{\alpha } \tan{\beta } - \tan{\beta } \tan{\gamma } - \tan{\alpha } \tan{\gamma }} ~,
 \label{eq:formula_tan_3}
\end{align}
we get
\begin{align}
 & \tan{\big(\tan^{-1}{\frac{(-2\mu ')}{m}} + \tan^{-1}{\frac{\mu ' + \sigma }{m  + \frac{1}{2}}} +   \tan^{-1}{\frac{\mu ' - \sigma }{m  + \frac{1}{2}}}\big)} = \frac{-(m+4)\mu '}{m^3 + m^2 +(4{\mu '}^2 + 8) m + 2 {\mu '}^2}~, 
 \label{eq:cal_tan} \\
 & \tan{\big(\tan^{-1}{2(\mu ' + \sigma )} +  \tan^{-1}{2(\mu ' - \sigma )}\big)} = \frac{\mu '}{8}~,
 \nonumber
\end{align}
Therefore, Eq. (\ref{eq:cal_log_0}) is finally given by
\begin{align}
 & \mu '\big(- \ln{\big(\frac{\epsilon }{2}\big)\big)} =  \lambda (\mu ') + \sum_{m=1}^{\infty } \tan^{-1}{\frac{(m+4)\mu '}{m^3 + m^2 +(4{\mu '}^2 + 2) m + 2 {\mu '}^2}}~,
  \label{eq:cal_log_0A} 
\end{align}
where
\begin{align}
 & \lambda (\mu ') = - \tan^{-1}\frac{\mu '}{8} + \tan^{-1}{\mu '} + \tan^{-1}{\frac{\mu '}{2}} +  \tan^{-1}{\frac{\mu '}{3}} +  \tan^{-1}{\frac{\mu '}{4}}
 \label{eq:cal_arctan} \\
 & = \tan^{-1}{\frac{7 \mu '}{8}}  + \tan^{-1}{\frac{\mu '}{2}} +  \tan^{-1}{\frac{\mu '}{3}} +  \tan^{-1}{\frac{\mu '}{4}} ~.
  \nonumber 
\end{align}
\indent Let us now assume that there exists a solution of $\mu ' > 0$. Then we can use an inequality $\tan^{-1}{x}< x$ for $x>0$ to yield
\begin{align}
 & \mu '\big(- \ln{\big(\frac{\epsilon }{2}\big)\big)} < \mu ' \big( \frac{7}{8+\mu '} + \frac{1}{2} + \frac{1}{3} + \frac{1}{4} \big) + \sum_{m=1}^{\infty } \frac{(m+4)\mu '}{m^3 + m^2 +(4{\mu '}^2 + 8) m + 2 {\mu '}^2}
  \label{eq:cal_mu_2} \\
 & < \frac{47}{24}\mu ' + \sum_{m=1}^{\infty }\frac{(m+4)\mu '}{m^3} = \mu ' \big( \frac{47}{24} + \zeta (2) + 4\zeta (3)\big) = \mu '\times 4.80~,
 \nonumber
\end{align}
where $\zeta (n)$ is the zeta function. Since $\epsilon \ll 1$, the above inequality does not hold. Therefore, there exist no solutions of $\mu ' >0$. \\
\indent For the case $\mu ' <0$, one can put as $\mu '=-\mu ''$, $\mu '' >0$. The Eq. (\ref{eq:condition_mu2a}) substituted $\mu '=-\mu ''$ corresponds to the complex conjugate equation of the original equation with $\mu '>0$. This fact shows that there exist also no solutions of $\mu '<0$.
 
%

\end{document}